\documentclass[sigconf]{acmart}
\AtBeginDocument{%
  }

\setcopyright{acmlicensed}
\copyrightyear{2018}
\acmYear{2018}
\acmDOI{XXXXXXX.XXXXXXX}
\acmConference[Conference acronym 'XX]{Make sure to enter the correct
  conference title from your rights confirmation email}{June 03--05,
  2018}{Woodstock, NY}
\acmISBN{978-1-4503-XXXX-X/2018/06}

\usepackage{enumitem}
\usepackage{xspace}
\usepackage{listings}
\usepackage{anyfontsize}
\usepackage[noend]{algpseudocode}
\usepackage{varwidth}
\usepackage{multirow}
\usepackage[table]{xcolor}
\usepackage[ruled,vlined]{algorithm2e}
\usepackage{subcaption}
\usepackage{array}

\usepackage{amssymb}
\usepackage{colortbl}
\newcolumntype{L}[1]{>{\raggedright\arraybackslash}p{#1}}

\newcommand{\name}{{\sf Halo}\xspace}

\begin{document}

\title{Batch Query Processing and Optimization for Agentic Workflows}

\author{Junyi Shen, Noppanat Wadlom, Yao Lu}
\affiliation{%
  \institution{National University of Singapore}
  \country{Singapore}
}
\email{{j1shen, noppanat, luyao}@comp.nus.edu.sg}


\begin{abstract}
Large Language Models (LLMs) in \textit{agentic workflows} combine multi-step reasoning, heterogeneous tool use, and collaboration across multiple specialized agents.
Existing LLM serving engines optimize individual calls in isolation, while multi-agent frameworks focus on orchestration without system-level performance planning.
As a result, repeated prompts, overlapping contexts, and fragmented CPU-GPU execution create substantial redundancy and poor hardware utilization, especially in batch analytics scenarios.

We introduce {\name}, a system that brings batch query processing and optimization into agentic LLM workflows.
{\name} represents each workflow as a structured query plan DAG and constructs a consolidated graph for batched queries that exposes shared computation.
Guided by a cost model that jointly considers heterogeneous resource constraints, prefill and decode costs, cache reuse, and GPU placement, {\name} performs plan-level optimization to minimize redundant execution.
The Processor integrates adaptive batching, KV-cache sharing and migration, along with fine-grained CPU-GPU pipelining to maximize holistic hardware efficiency.

Evaluation across six benchmarks shows that \name{} achieves up to 3.6$\times$ speedup for batch inference and 2.6$\times$ throughput improvement under online serving, scaling to workloads of thousands of queries and complex graphs.
These gains are achieved without compromising output quality.
By unifying query optimization with heterogeneous LLM serving, \name{} enables efficient agentic workflows in data analytics and decision-making applications.
\end{abstract}

\received{20 February 2007}
\received[revised]{12 March 2009}
\received[accepted]{5 June 2009}

\maketitle

\section{Introduction} 
\begin{figure}
    \centering
    \includegraphics[width=1\linewidth]{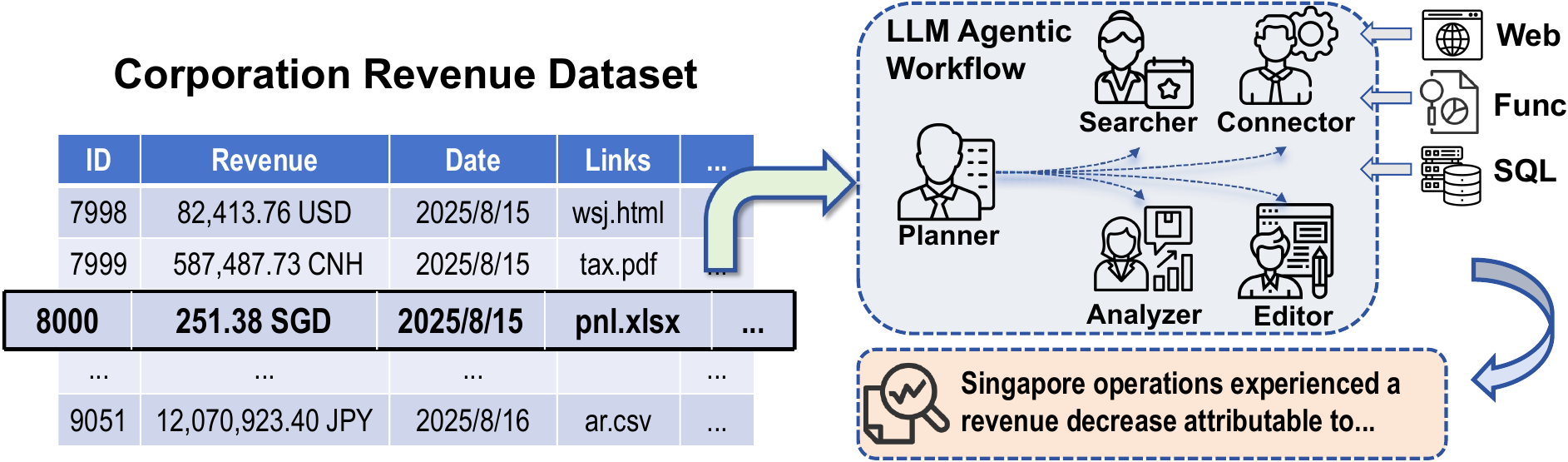}
    \caption{\small{An example agentic workflow in which multiple LLM agents collaborate to analyze revenue data and provide decision support for businesses.}}
    \label{fig:intro}
\end{figure}

Large Language Models (LLMs) and agentic workflows~\cite{gao2025beyond, jin2025search} are increasingly employed to address complex tasks in data analytics and decision making. An \emph{agentic LLM workflow} is a goal-driven sequence of operations involving multiple LLM invocations, potentially with different models, prompts, and roles, that collaborate to produce a final result. Such workflows typically require multi-step reasoning, tool use, pipelined execution, and are often orchestrated by human users or specialized planning LLMs~\cite{wei2022chain,lewis2020retrieval,yao2022react,schick2023toolformer}. Recent work has further explored multi-agent composition and simulations to model sophisticated scenarios (e.g., negotiation, or market dynamics), which amplifies both structural complexity and computational demand~\cite{wu2024autogen,gao2025agentscope,park2024generative,langgraph}.
Consider a data analytics assistant that investigates anomalous revenue drops (Figure~\ref{fig:intro}). A lead \emph{planner} delegates to: a \textit{searcher} to retrieve aggregated data, an \textit{analyzer} to run attribution, a \textit{connector} to correlate product events, incident tickets, and release notes, and an \textit{editor} to synthesize hypotheses with evidence and counterfactual checks. The workflow fans out into many retrieval-and-summarize calls over overlapping contexts, repeatedly drafts/refines common SQL queries (e.g., weekly revenue by SKU) and API calls (e.g., fetching and parsing news), and then aggregates results. At scale, tens to hundreds of investigations may run concurrently, with frequent LLM calls, prompt reuse, and intermediate summaries. 
In practice, batch analytics queries are common~\cite{zheng2023judging,hou2024bridging, zhuge2024agent}: the same workflow may be replicated across different markets, product lines, or time frames, reusing query templates and agentic workflow structures but contextualized with different data insights.

Recent advances in LLM serving engines have substantially reduced per-request latency via speculative decoding and optimized attention kernels~\cite{fastertransformer,kwon2023efficient,leviathan2023fast,li2025eagle,zheng2024sglang}. 
However, they largely optimize each workflow DAG in isolation and ignore heterogeneous CPU-side tool operators (e.g., SQL/HTTP/parsing). 
Meanwhile, multi-agent frameworks (LangGraph, AutoGen, AgentScope) ease workflow construction but typically leave performance to downstream serving/orchestration rather than plan-level optimization~\cite{langgraph,wu2024autogen,gao2025agentscope}. 
General DAG data systems such as Dask and Spark can execute DAG plans, yet treat operators as black boxes and thus cannot exploit LLM-specific opportunities like adaptive prefill/decode scheduling or KV-cache reuse~\cite{moritz2018ray,rocklin2015dask,zaharia2010spark}. 
Finally, heterogeneity-aware schedulers such as \emph{Gavel}~\cite{narayanan2020heterogeneity} and \emph{Hadar}~\cite{sultana2025resource} focus on multi-GPU cluster scheduling for \emph{atomic} deep-learning training jobs, optimizing allocation across heterogeneous accelerators rather than jointly optimizing mixed CPU--GPU operators \emph{within} agentic DAG batches.
Consequently, batched agentic workflows remain inefficient with off-the-shelf solutions, motivating a runtime that performs cross-DAG and heterogeneous CPU--GPU co-optimization under mixed LLM and tool-centric operators.

In this paper, we present \name{}\footnote{Our source code is at \url{https://anonymous.4open.science/r/Halo_Demo-BC86/}.}, a system that extends batch query processing and optimization to heterogeneous agentic workflows in data analytics and decision-making. In contrast to existing serving engines and agent frameworks that optimize each query or workflow in isolation, Halo models each workflow as a structured query-plan DAG, whose nodes correspond to GPU-resident LLM invocations or CPU-executed external function calls. This representation enables Halo to draw on decades of database optimization principles while explicitly accounting for the distinct execution characteristics of LLM inference and tool-mediated operators.

\name{}’s query processor and optimizer build on two core insights. First, instead of relying on rule-based graph rewrites or Volcano-style optimization pipelines~\cite{graefe1993volcano}, \name{} formulates optimization and execution as a dependency-aware placement and scheduling problem over heterogeneous workers. Each workflow is compiled into a structured DAG with explicit precedence constraints; operators include GPU-resident LLM prefill/decode stages as well as CPU-bounded tool operators (e.g., database queries, retrieval, parsing, and API calls) whose runtimes and contention characteristics differ substantially from inference. Halo then selects placements and schedules guided by a cost model that captures (i) prefill–decode interactions, (ii) KV-cache reuse and sharing across common prefixes, and (iii) CPU-side tool contention and communication overheads. This direct optimization both simplifies the system architecture and yields near-optimal performance in practice. Second, \name{} integrates state-of-the-art LLM serving techniques, including adaptive batching, cache sharing/prefetching, and overlap of compute and communication, within a unified DAG scheduler that respects inter-operator dependencies and avoids pathological stalls (e.g., GPU underutilization while CPU tools are backlogged, or tool results delaying decode fan-in). 

Returning to the revenue investigation example, \name{} automatically detects recurring prompts, batches overlapping summarization requests across concurrent agents, reuses KV caches for shared prefixes, and pipelines prefill and decode across multiple GPU workers while concurrently scheduling CPU-bound tool operators under dependency and resource constraints. This holistic, plan-level optimization improves end-to-end latency and throughput.

We implement \name{} to integrate seamlessly with existing LLM backends (e.g., vLLM) to accommodate both offline batch queries and online streaming queries. Across six benchmark workloads, \name{} delivers up to {3.6×} batch-inference speedups over state-of-the-art agentic frameworks (and over 400× against commonly-used engines). For online serving, \name{} offers up to {2.6×} throughput improvements. We further validate \name{}’s robustness and flexibility under scaled workloads, featuring high-volume queries (up to 2K) and complex topologies, across heterogeneous model sizes (up to 32B) and hardware generations.

To summarize, the contributions of this work are as follows:
\begin{itemize}[leftmargin=0.15in]
  \item \emph{A novel system} that combines LLM serving with query optimization for heterogeneous workflows (GPU LLM and CPU-bounded tool operators); to our knowledge, it is the first to do so.
  \item \emph{A dependency- and resource-aware optimizer} for precedence-constrained placement and scheduling over heterogeneous workers, and a dynamic-programming solver for optimality. 
  \item \emph{A query processor} that enables adaptive cross-workflow batching, KV-cache sharing/prefetching, and compute–communication overlap while respecting DAG dependencies.
  \item {Comprehensive evaluations} show significant speedups in latency and throughput for both offline batch and online streaming queries without degrading output quality.
\end{itemize}

\section{Background and Motivation}
\label{sec:background}


We focus on \textit{workflow-based agents}, a paradigm widely used in enterprise applications to enforce structured reasoning or standard operating procedures~\cite{chase2022langchain, khattab2023dspy}.
We model such an application as a DAG $G = (V, E)$, where each node $v \in V$ is a schedulable unit of computation and each edge $(u, v) \in E$ encodes a data or control dependency~\cite{shen2023hugginggpt, besta2024graph}.
Unlike open-ended autonomous loops, a user request here expands into a structured DAG execution with a static topology.

\vspace{0.05in}\noindent\textbf{Nodes with heterogeneous resource usage.}
The nodes in agentic workflows fall into two primary categories:
(i) \emph{LLM nodes}, which invoke an LLM on a prompt to produce textual outputs, usually with a GPU runtime; and
(ii) \emph{Tool usage nodes}, which execute generic computational or retrieval tasks via function calls, usually with a CPU runtime~\cite{schick2023toolformer, yao2022react}.
This broad category may contain structured database queries (SQL), external APIs (e.g., HTTP requests), and local function executions.
Each node $v$ is associated with a type
$t(v) \in \{\text{LLM}, \text{Tool}\}$ and a configuration
(e.g., decoding parameters for LLMs, or function signatures for tools).

\begin{table}[t]
  \centering
  \small 
  \begin{tabular}{l c}
    \toprule
    Research Project & Number of Agents \\
    \midrule
    Generative Agents~\cite{park2023generative} & 25 \\
    \rowcolor{gray!10}
    Project Sid~\cite{al2024project} & 500 \\
    AI Metropolis~\cite{xie2025ai} & 1,000 \\
    \rowcolor{gray!10}
    Generative Agent Simulations~\cite{park2024generative} & 1,052 \\
    LMAgent~\cite{liu2024lmagent} & 10,000 \\
    \bottomrule
  \end{tabular}\vspace{0.05in}
  \caption{The increasing scale of multi-agent research projects creates massive concurrency, necessitating efficient batching.}
  \label{tab:AgentNum}
\end{table}

\begin{figure}[t]
    \centering
    \includegraphics[width=0.9\linewidth]{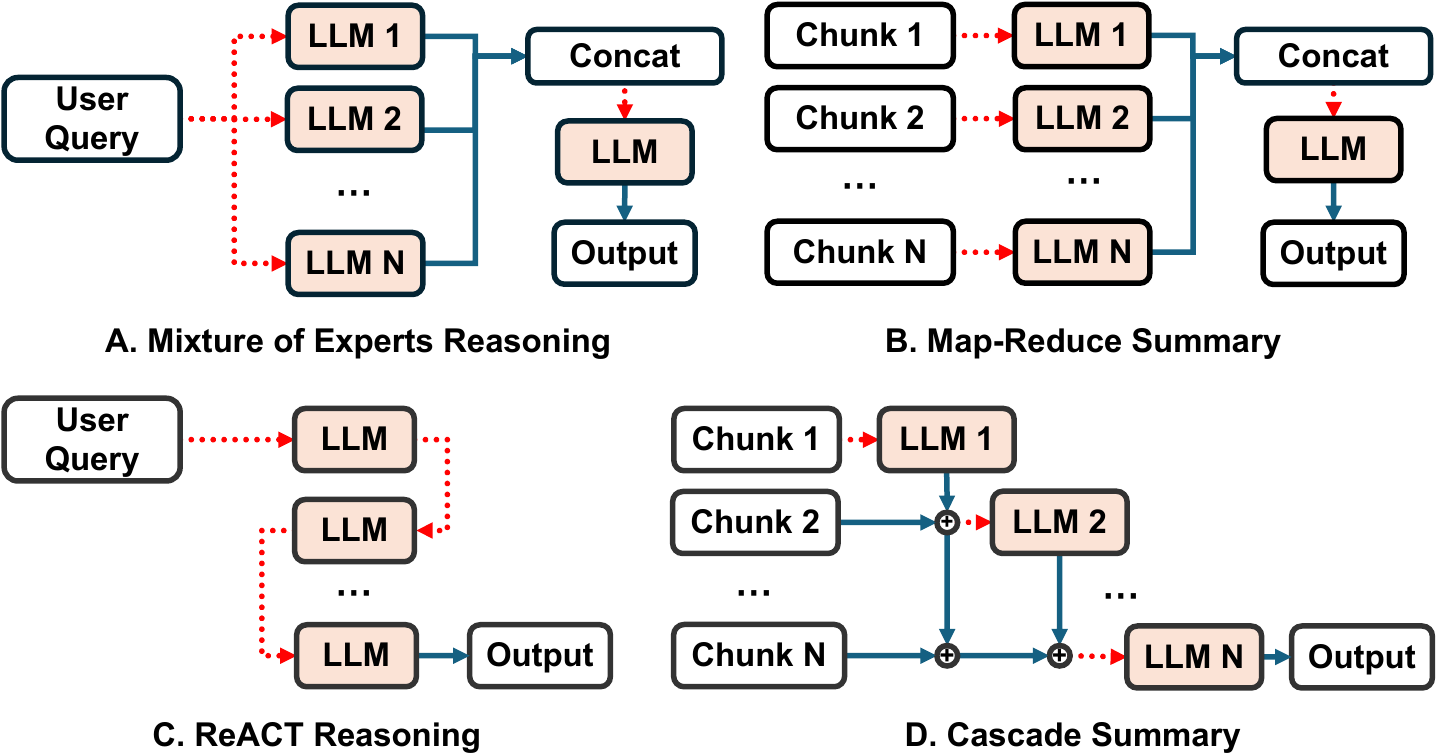}\vspace{-0.05in}
    \caption{Complex agentic workflows involve modular, collaborative, and adaptive processes.}
    \label{fig:workflows}
\end{figure}

\vspace{0.05in}\noindent\textbf{Heterogeneous job scheduling: opportunities \& challenges.} The rapid scaling of multi-agent systems requires a throughput solution. As summarized in Table~\ref{tab:AgentNum}, recent simulations already involve thousands of concurrent agents (e.g., up to 10,000 in LMAgent~\cite{liu2024lmagent}). This level of concurrency is not merely a stressor; it also unlocks optimization headroom that is largely absent in single-request settings. Moreover, real-world agentic workloads are often non-linear; they form structured DAGs with fan-out, fan-in, and iterative loops~\cite{wu2024autogen, qian2024chatdev, lin2024parrot}, as illustrated in Figure~\ref{fig:workflows}. Such a structure creates a strong \emph{locality} that a global scheduler can exploit:
\begin{itemize}[leftmargin=10pt, topsep=2pt, itemsep=0pt]
\item \emph{Model reuse:} Parallel branches often invoke the same model (e.g., multiple refinement steps using Qwen-32B). Co-locating and batching these invocations reduces repeated model (re-)loading and improves GPU residency.
\item \emph{Context reuse:} Child nodes typically extend parent contexts. Preserving lineage on the same worker enables prefix caching and KV reuse~\cite{zheng2024sglang}, avoiding redundant prefill over shared history.
\item \emph{Tool coalescing:} Many workflows issue identical or overlapping tool calls (e.g., SQL retrieval queries) across agents and sessions. Coalescing these requests to a single physical execution amortizes overhead and enables shared reuse of results~\cite{facebook2019dataloader,giannikis2012shareddb}.
\end{itemize}

These opportunities manifest differently across heterogeneous operators. For GPU-resident LLM inference, performance is largely constrained by memory bandwidth, and multi-model workflows additionally incur substantial weight-loading overhead. Continuous batching amortizes weight access by executing many concurrent requests in a single forward pass~\cite{yu2022orca}, while high concurrency enables intra-batch prefix sharing that computes shared prefixes once and reuses the resulting KV cache across requests~\cite{zheng2024sglang}. In contrast, DB and tool operators are often CPU- and I/O-bound; here the primary levers are request coalescing (read deduplication)~\cite{facebook2019dataloader} and locality-aware scheduling. Merging identical concurrent requests, which can be common when multiple agents query the same “world state”, avoids redundant executions; this is akin to multiplexing in shared query workloads~\cite{giannikis2012shareddb}. Routing similar requests to the same worker further exploits warm state (e.g., prepared statements and buffer pools), reducing parsing overhead and minimizing disk I/O~\cite{hellerstein2007architecture}.

\vspace{0.05in}\noindent\textbf{Challenges}. Given a workflow DAG and a set of heterogeneous workers, scheduling agentic workflows is markedly harder than conventional request-level LLM serving or single-query DB optimization. We highlight three fundamental challenges below:

\begin{itemize}[leftmargin=*]
    \item \emph{C1: Structural awareness.} Agentic workflows exhibit significant {structural redundancy}, where multiple agents often traverse identical subgraphs (e.g., shared SOPs). Existing schedulers execute these requests as independent events, causing {redundant I/O and computation}. A workflow-aware scheduler must identify these global patterns to enable {request coalescing}, rather than optimizing individual nodes in isolation.

    \item \emph{C2: Heterogeneous pipeline bubbles.} Execution alternates between CPU-bound tool usage or database calls and GPU-bound LLM inference. Prior DAG-based machine learning schedulers ~\cite{topcuoglu2002performance, moritz2018ray} often overlook this, leading to \emph{Pipeline Bubbles} where GPUs sit idle waiting for context. Minimizing latency requires precise orchestration of the heterogeneous pipeline.

    \item \emph{C3: Stateful LLM operators.} Execution cost fluctuates wildly based on worker state. Switching models incurs expensive weight loading costs (\textit{model state}), while missing warm KV-caches or tool connections increases inference latency (\textit{context}). 
\end{itemize}

\begin{table}[t]
  \centering
  \small
  \caption{\small{Comparing existing solutions. \textbf{C1}: Structure Awareness (Global DAG); \textbf{C2}: Heterogeneity (CPU-GPU Runtime); \textbf{C3}: Stateful Optimization.}}
  \label{tab:comparison}
  \begin{tabular}{l l c c c}
    \toprule
    \textbf{Solution} & \textbf{Examples} & \textbf{C1} & \textbf{C2} & \textbf{C3} \\
    \midrule
    Request-level serving & \emph{vLLM}, \emph{SGLang} & $\times$ & $\times$ & $\sim$\textsuperscript{\dag} \\
    Data platforms & \emph{Ray}, \emph{Spark} & $\sim$\textsuperscript{\ddag} & $\times$ & $\times$ \\
    App orchestrators & \emph{LangGraph}, \emph{AgentScope} & \checkmark & $\times$ & $\sim$\textsuperscript{\S} \\
    \midrule
    \rowcolor{gray!10}
    \textbf{Halo (Our solution)} & - & \textbf{\checkmark} & \textbf{\checkmark} & \textbf{\checkmark} \\
    \bottomrule
    \multicolumn{5}{l}{\footnotesize \textsuperscript{\dag} \emph{Optimizes internal KV-cache but lacks cross-request logic awareness.}} \\
    \multicolumn{5}{l}{\footnotesize \textsuperscript{\ddag} \emph{Supports DAG execution but treats tasks as opaque black boxes.}} \\
    \multicolumn{5}{l}{\footnotesize \textsuperscript{\S} \emph{State is external to the runtime, preventing proactive intervention.}}
  \end{tabular}
\end{table}

\label{sec:background-challenges}

Table~\ref{tab:comparison} compares off-the-shelf solutions in handling these challenges: (1) \emph{LLM serving engines} such as \emph{vLLM}, \emph{SGlang}, and \emph{TGI}~\cite{kwon2023efficient, tgi, zheng2024sglang} deliver efficient continuous batching on GPUs, but treat requests as independent and thus miss logical deduplication; (2) \emph{general-purpose data systems} with worker schedulers such as \emph{Ray} ~\cite{moritz2018ray} allocate resources coarsely but are largely \emph{workflow-agnostic}; they do not exploit DAG dependencies for co-location or state reuse;  (3) \emph{application-level orchestrators} such as \emph{LangGraph} and \emph{AgentScope}~\cite{langgraph, gao2402agentscope} capture agent interaction logic but are decoupled from the serving runtime. Without a global view across sessions, they cannot perform {cross-session batching}, fragmenting mergeable work into separate API calls and limiting scalability under concurrency. Moreover, all these solutions poorly handle coupled CPU tool phases and GPU inference, often leaving GPUs idle during CPU-bound execution. 

\section{System Overview}
\label{sec:overview}
\begin{figure}
    \centering
    \includegraphics[width=1\linewidth]{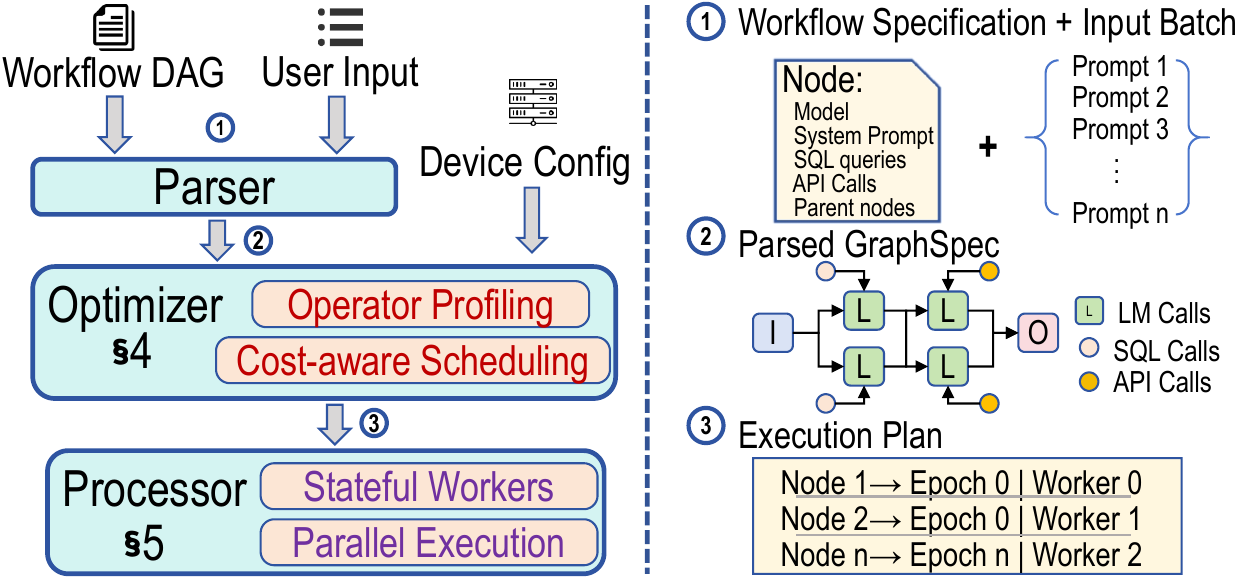}
    \caption{\name system overview.}
    \label{fig:arch}
\end{figure}

In this paper, we aim for a holistic optimization to handle these challenges. Batch agent workflow DAGs are scheduled among heterogeneous CPU-GPU workers jointly rather than optimizing each request in isolation.
Figure~\ref{fig:arch} illustrates the architecture of \name, with a classic parser-optimizer-processor architecture:

\smallskip
\noindent\textbf{Parser.}
The Parser ingests a declarative workflow in YAML and converts it into a typed intermediate representation~\cite{lattner2020mlircompilerinfrastructureend} (\texttt{GraphSpec}).
A key transformation is {dependency decoupling} to extract embedded non-LLM computations, such as database queries, external API invocations, and local helper functions, into standalone nodes. This normalization ensures that the scheduler treats these CPU-bound operations as schedulable units, rather than opaque side-effects inside LLM prompts; this enables the joint optimization strategies discussed in Section~\ref{sec:background-challenges}.

\smallskip \noindent\textbf{Optimizer.} The Optimizer takes a \texttt{GraphSpec} and produces an \texttt{ExecutionPlan} through a two-phase process. First, the \emph{Operator Profiler} evaluates all nodes within the workflow, covering both LLM and non-LLM operators~\cite{marcus2019neo, luo2025autellix}. It analyzes these operators to estimate resource consumption and latency, establishing a comprehensive data-driven cost foundation~\cite{selinger1979access} to support the subsequent orchestration. Second, the \emph{Solver} leverages these profiles to run a global optimization algorithm (detailed in Section~\ref{sec:optimizer}). Unlike conventional optimizers that assume fixed costs, \name's solver jointly decides (i) placements of LLM nodes to maximize context reuse, (ii) execution strategies for non-LLM operators, and (iii) precise task orderings to balance CPU-GPU load and form efficient batches.

\smallskip
\noindent\textbf{Processor.}
The processor serves as the runtime engine responsible for coordinating a pool of long-running workers, including GPU workers that host vLLM
instances for LLM inference and CPU threads that execute tool usage tasks (e.g., database queries, external API calls, and local functions).
It executes a given \texttt{ExecutionPlan} by orchestrating task dispatch, batching, and result collection across these heterogeneous workers.
To minimize runtime overhead, the Processor exploits dynamic {pipeline parallelism}~\cite{narayanan2019pipedream} to create distinct execution streams where GPU and CPU nodes run concurrently whenever data dependencies permit~\cite{augonnet2009starpu, bauer2012legion}.

\vspace{0.05in}\noindent\textbf{Scope}.
 We note the following scope for our solution:
\begin{itemize}[leftmargin=0.15in]
    \item \emph{Semantic preserving.} We require strict equivalence to the original execution; no approximate techniques (e.g., proxy models) or prompt modifications are used.
    \item \emph{Single-machine, multi-GPU.} We focus on a single server setting to enable fine-grained, white-box control over GPU memory, batching, and data locality. 
    \item \emph{Model \& tool usage.} We aim for workflows that invoke diverse LLM models interspersed with CPU-intensive tool calls. Consequently, the scheduler must actively manage model weights to mitigate switching penalties, while simultaneously orchestrating CPU resources to prevent tool execution from bottlenecking the inference pipeline.
\end{itemize}
\section{Optimizer Design}
\label{sec:optimizer}

We formulate \name's optimizer as a {GPU-centric}, resource-aware placement and scheduling problem over {agentic} LLM workflows. While the runtime executes a heterogeneous graph with both GPU-resident LLM operators and CPU-bound tool operators (e.g., SQL, HTTP, parsing), the optimizer's {explicit} decision variables focus on scarce GPU workers. CPU-side tool execution is handled via an implicit orchestration policy and is incorporated into the objective through profiled preparation costs.

\vspace{0.05in}\noindent\textbf{Epoch-based scheduling.}
Directly optimizing continuous start times yields an intractable search space for online planning. Following prior work on discretized scheduling~\cite{ullman1975np, kwok1999static}, we partition planning into decision windows (i.e., \emph{epochs})~\cite{zaharia2013discretized, grandl2014multi, kannan2019grandslam}. At the beginning of epoch $e$, the optimizer leverages the current system state and selects a set of LLM operators to launch on GPU workers; the Processor (Sec.~\ref{sec:processor}) then realizes the plan with finer-grained pipelining and opportunistic execution at runtime.

Let $G_{\text{LLM}}=(V,E)$ denote the dependency DAG over LLM operators. The system state at the start of epoch $e$ is
\[
S_e = (D_e, H_e).
\]
Here, $D_e \subseteq V$ is the set of completed LLM nodes, and 
\[
H_e=\{h_w^e \mid w\in W\}
\] 
summarizes the persistent GPU worker context,
where $h_w^e$ captures (i) the resident model weights on worker $w$ and (ii) a compact representation of its warm KV-cache state, both
of which strongly influence the latency of subsequent LLM operators. Executing an epoch action updates the completed set
($D_{e+1}=D_e\cup B_e$) and deterministically transitions worker contexts to $H_{e+1}$,  where $B_e$ are the nodes picked in the current epoch.

\vspace{0.05in}\noindent\textbf{Objective.}
Given a finite sequence of actions (state transitions) $\mathcal{A}=\{A_0,A_1,\dots,A_{E-1}\}$ in epochs that reach completion, Halo minimizes the total
epoch cost:
\[
\min_{\mathcal{A}} \ \sum_{e=0}^{E-1} C_{\text{epoch}}(S_e, A_e),
\]
where $E$ is the number of epochs until all LLM nodes complete.
An epoch action $A_e$ induces an estimated per-worker time budget. Let $\phi_e(v)$ be the worker assigned to node $v$, and let
\[
T_w \;=\; \sum_{v \in B_e:\ \phi_e(v)=w} T(w,v,S_e),
\]
where $T(w,v,S_e)$ is the profiled, state-aware latency estimate for executing $v$ on worker $w$ under state $S_e$ (Section~\S\ref{sec:optimizer-cost}). We define the epoch cost as
\[
C_{\text{epoch}}(S_e,A_e) \;=\; \mu \cdot \max_{w\in W} T_w \;+\; (1-\mu)\cdot \sum_{w\in W} T_w \;+\; \lambda \cdot g(A_e).
\]
The $\max$ term targets makespan reduction by minimizing the bottleneck worker. The $\sum$ term penalizes aggregate system load,
preventing the optimizer from ignoring inefficiencies off the critical path. Finally, $g(A_e)$ regularizes fixed per-epoch overhead
(e.g., dispatch and coordination), discouraging excessive fragmentation of the plan.

\smallskip
\noindent\textbf{Action space.}
In epoch $e$, the optimizer selects a \emph{scheduling policy} $A_e=(B_e,\phi_e)$, with
\begin{itemize} [leftmargin=0.15in]
  \item $B_e \subseteq V\setminus D_e$: the set of LLM nodes launched in epoch $e$, and
  \item $\phi_e: B_e \rightarrow W$: the mapping from selected nodes to GPU workers.
\end{itemize}
This design makes GPU assignment explicit, reflecting its role as the dominant constraint in multi-model, multi-agent serving, while
allowing CPU-side work to overlap via pipeline parallelism at runtime.
Here, a decision $A_e$ is feasible if it respects (1) \emph{Precedence constraints.} Nodes selected in $B_e$ must form a valid topological cut: for each $v\in B_e$, every direct   LLM predecessor of $v$ must belong to $D_e$ or to $B_e$ (so that dependent LLM steps may be chained within the same epoch and   executed sequentially by the runtime in dependency order). (2) \emph{Resource constraints.} Each GPU worker can host at most one \emph{concurrent} LLM execution at a time; to keep the search space tractable in a single-machine setting, we constrain $\phi_e$ to be injective (i.e., $|B_e|\le |W|$ and no worker is assigned two nodes \emph{as primary epoch launches}).
Within an epoch, the Processor may still exploit fine-grained pipelining and batching, but the optimizer's abstraction ensures a
physically consistent and dependency-respecting launch set.

Given a fixed GPU schedule $A_e$, prerequisite CPU tool operators required to supply inputs for the selected LLM nodes are automatically inferred and dispatched by the Processor. Concretely, \name prioritizes CPU tasks by DAG depth (shallower dependencies first), which tends to resolve critical-path prerequisites early and improves CPU--GPU overlap, akin to rank-based list scheduling heuristics. Their contribution is reflected in $T(w,v,S_e)$ through the profiled preparation component (Section~\S\ref{sec:optimizer-cost}), enabling the optimizer to reserve sufficient lead time to avoid GPU starvation.

\smallskip
\noindent\textbf{Dynamic programming solver.}
Previously, exact continuous-time formulations (e.g., MILP) target global optimality but face the NP-complete barrier~\cite{ullman1975np}, preventing online use.
Conversely, greedy heuristics like HEFT~\cite{topcuoglu2002performance} or SRPT~\cite{bansal2001analysis} are fast but myopic, ignoring hidden costs like CPU context preparation.

\name bridges this gap via {discretized approximation} while ensuring optimality. By pruning the state space to topological frontiers, we reduce complexity from exponential to near-linear in practice, which is orders of magnitude faster than MILP. 
This makes global planning tractable while retaining the "white-box" awareness required for efficient pipeline orchestration. 

\begin{algorithm}[t]
\caption{Halo: Holistic Scheduling for Heterogeneous LLM Workflows}
\label{alg:dp-scheduler}
\small 

\SetAlgoLined
\SetAlgoNoEnd
\DontPrintSemicolon
\SetArgSty{textnormal}
\SetKwInOut{Input}{Input}
\SetKwInOut{Output}{Output}
\newcommand\myCommentStyle[1]{\footnotesize\ttfamily\textcolor{gray}{#1}}
\SetCommentSty{myCommentStyle}
\SetKwComment{Comment}{// }{}
\SetKwProg{Func}{Function}{}{}
\let\oldnl\nl
\newcommand{\nonl}{\renewcommand{\nl}{\let\nl\oldnl}}

\SetKwFunction{GetFrontier}{GetFrontier}
\SetKwFunction{EstCost}{CostModel}
\SetKwFunction{UpdateH}{UpdateWorkerState}

\Input{DAG $G=(V,E)$; Workers $\mathcal{W}$; Initial $H^0$}
\Output{Optimal Policy $\Pi^{\star}$}
\textbf{Global:} Table $\mathsf{M} \gets \varnothing$; $c^{\star} \gets \infty$\;

\smallskip
\Func{\textsc{Solve}($D, H, e$)}{
    \lIf{$D = V$}{ \Return $(0,\ \varnothing)$ \Comment*[f]{Termination}}
    
    \lIf{$(D,H) \in \mathsf{M}$}{ \Return $\mathsf{M}[(D,H)]$ \Comment*[f]{Memo hit}}

    $\mathcal{F} \gets \GetFrontier(G, D)$\Comment*[f]{Topological ready set}\;

    \ForEach(\Comment*[f]{Search}){batch $B \subseteq \mathcal{F}$ \textbf{and} map $\phi: B \to \mathcal{W}$}{
        $c_{\text{now}} \gets \EstCost(B, \phi, H, \widehat{T}_{\text{prep}}, \widehat{T}_{\text{model}}, \widehat{T}_{\text{infer}})$\;
        $H_{\text{next}} \gets \UpdateH(H, \phi)$ \Comment*[r]{Update State}
        $(c_{\text{fut}}, \Pi_{\text{rest}}) \gets \textsc{Solve}(D \cup B, H_{\text{next}}, e{+}1)$\;
        
        \lIf{$c_{\text{now}} + c_{\text{fut}} < c^{\star}$}{
            $c^{\star} \gets c_{\text{now}} + c_{\text{fut}}$; $\Pi^{\star} \gets \{(B, \phi)\} \cup \Pi_{\text{rest}}$
        }
    }
    
    $\mathsf{M}[(D,H)] \gets (c^{\star}, \Pi^{\star})$\;
    \Return $(c^{\star}, \Pi^{\star})$\;
}

\textbf{Main:} \Return \textsc{Solve}$(\varnothing, H^{0}, 0)$\Comment*[f]{Trigger global planning}\;
\end{algorithm}

Specifically, we adopt an epoch-based DP that explicitly captures stateful GPU effects (resident weights and KV-cache) in $H_e$, satisfying a Markov property: the future cost depends on the past only through the current $(D_e,H_e)$. Let $V(S)$ be the minimum cumulative cost from state $S$ to completion. The Bellman recurrence is:
\[
V(S) \;=\; \min_{A\in \mathcal{A}(S)} \Bigl( C_{\text{epoch}}(S,A) \;+\; V(\textsc{Next}(S,A)) \Bigr),
\]
where $\mathcal{A}(S)$ enumerates feasible epoch actions reachable from $S$ (restricted to valid topological frontiers), and
$\textsc{Next}(S,A)$ applies the deterministic updates to $(D,H)$. 

Algorithm~\ref{alg:dp-scheduler} implements this recurrence with
memoized recursion: it enumerates frontier batches and worker assignments, evaluates each candidate using the cost model, simulates the
worker-state transition, and reuses cached subproblem solutions to avoid redundant exploration. The core function \textsc{Solve} explores the state space defined by the completed set $D$ and hardware context $H$, using a hash table $\mathsf{M}$ to cache optimal sub-paths and prevent redundant computations.
Crucially, the search is guided by \texttt{GetFrontier}, which restricts candidate actions to valid topological cuts, thereby enforcing the dependency constraints.
For each candidate batch, \texttt{CostModel} evaluates the epoch cost $C_{\text{epoch}}$ based on the profiling metrics derived in Section~\ref{sec:optimizer-cost}, while \texttt{UpdateWorkerState} simulates the deterministic transition of GPU states (e.g., model loading and cache eviction), ensuring the plan remains physically consistent.

\smallskip
\noindent\textbf{Hardness and complexity analysis.} Let $N_V=|V|$ be the number of LLM nodes, $M=|W|$ the number of GPU workers, and $N_m$ the number of distinct models. In the worst case, the DP state space is exponential in $N_V$ due to the completed-set component ($2^{N_V}$), and multiplicative in the number of possible per-worker contexts (dominated by model residency and cache identifiers), yielding a upper bound of
$O\!\left(2^{N_V}\cdot (N_m N_V)^M\right)$. In practice, the reachable subset is far smaller: valid states are constrained by the DAG's
topological structure and grow primarily with the maximum frontier width rather than the total length. Since single-machine settings
typically have small $M$ (e.g., $\le 8$ GPUs), and frontier widths are bounded by workflow parallelism, the resulting search remains
tractable while retaining the ``white-box'' state awareness necessary for model-switch and KV-cache optimization.



\subsection{Cost Modeling}
\label{sec:optimizer-cost}

\name's solver relies on a state-aware latency estimator \(T(w,v,S_e)\) to evaluate candidate epoch actions. The estimator is designed to (i) reflect the heterogeneous CPU--GPU nature of agentic workflows, and (ii) capture the dominant \emph{stateful} effects on GPUs, including model residency and KV-cache reuse.

\noindent \textbf{State-aware latency estimation ($T$).} 
For an LLM operator \(v\) assigned to GPU worker \(w\) at epoch \(e\), we model end-to-end latency as the additive sum of (a) prerequisite CPU-side preparation and (b) GPU-side execution conditioned on the worker state \(h_w^e\):
\begin{equation}
\label{eq:latency_decomp}
T(w,v,S_e) \;=\; T_{\text{prep}}(v) \;+\; T_{\text{gpu}}(w,v,h_w^e).
\end{equation}
The preparation term \(T_{\text{prep}}(v)\) accounts for upstream tool operators (e.g., SQL/HTTP/local functions) and data materialization required to make \(v\) runnable on the GPU. The GPU term depends on the persistent worker context \(h_w^e\), which we summarize as \(h_w^e=(m_w^e,u_w^e)\), where \(m_w^e\) is the resident model weights and \(u_w^e\) is a compact signature of the worker's warm KV-cache state.
 
We further decompose \(T_{\text{gpu}}\) into a \emph{model-switch} component and an \emph{inference}
component:
\begin{equation}
\label{eq:gpu_cost}
T_{\text{gpu}}(w,v,h_w^e)
\;=\;
T_{\text{model}}(v, m_w^e)
\;+\;
T_{\text{infer}}(v, u_w^e).
\end{equation}
Multi-model agentic workflows incur substantial overhead when a worker must load weights for a model
different from its currently resident \(m_w^e\). We represent this effect as
\(T_{\text{model}}(v,m_w^e)=0\) when \(m_w^e\) matches the model required by \(v\), and a profiled
reload penalty otherwise (including any eviction/memory management cost required to admit the new
model). This encourages the solver to schedule model-homogeneous runs and to avoid pathological thrashing.

We note that inference latency depends on the prompt length, generation length, and whether the worker can reuse
KV-cache entries from a previously executed lineage. We model \(T_{\text{infer}}(v,u_w^e)\) as a
state-conditional estimator that accounts for (i) prefill work and (ii) decode work. When the prompt of
\(v\) shares a cached prefix with \(u_w^e\), the estimator applies a \emph{prefix-caching discount} by
reducing the effective prefill tokens; otherwise it charges full prefill. This term drives locality-aware
assignment by rewarding schedules that preserve workflow lineage on the same worker.

As a result, Equation~\eqref{eq:latency_decomp} makes CPU-side preparation explicit. By charging
\(T_{\text{prep}}(v)\) in \(T(w,v,S_e)\), the optimizer internalizes the risk of launching GPU work before
its prerequisite tool outputs are ready. This design reduces \emph{pipeline bubbles} (GPU starvation
due to tool backlog) because candidate epoch actions that overcommit GPUs without sufficient
preparation time become immediately expensive under the objective.


\smallskip
\noindent\textbf{Operator profiling and online calibration.}
\name uses lightweight profiling to instantiate the three components
\(T_{\text{prep}}\), \(T_{\text{model}}\), and \(T_{\text{infer}}\) without imposing material planning overhead.
\begin{itemize}[leftmargin=0.15in]
  \item \emph{Database operators.} For SQL preparation, Halo queries the DBMS \texttt{EXPLAIN} interface to obtain cost estimates for candidate physical plans (e.g., index scan vs.\ sequential scan), and uses this information to predict execution time under the current schema and statistics.
  \item \emph{Black-box tools and external APIs.} For HTTP requests and user-defined functions, \name maintains a moving average (and optionally a bounded-variance estimate) over recent invocations, keyed by a normalized signature of the operator and its arguments.
  \item \emph{LLM inference.} For each model and hardware configuration, Halo maintains calibrated throughput curves that map (effective) prefill tokens and decode tokens to latency. Prefix reuse is reflected by reducing effective prefill tokens according to the matched cached-prefix length implied
  by \(u_w^e\).
\end{itemize}
Lastly, profiling statistics are updated continuously, allowing the processor to refine cost estimates as runtime conditions drift (e.g., due to contention, network jitter, or decoding variance), while preserving the solver's epoch-level abstraction.


\section{Processor Design}
\label{sec:processor}
\begin{figure}
    \centering
    \includegraphics[width=1\linewidth]{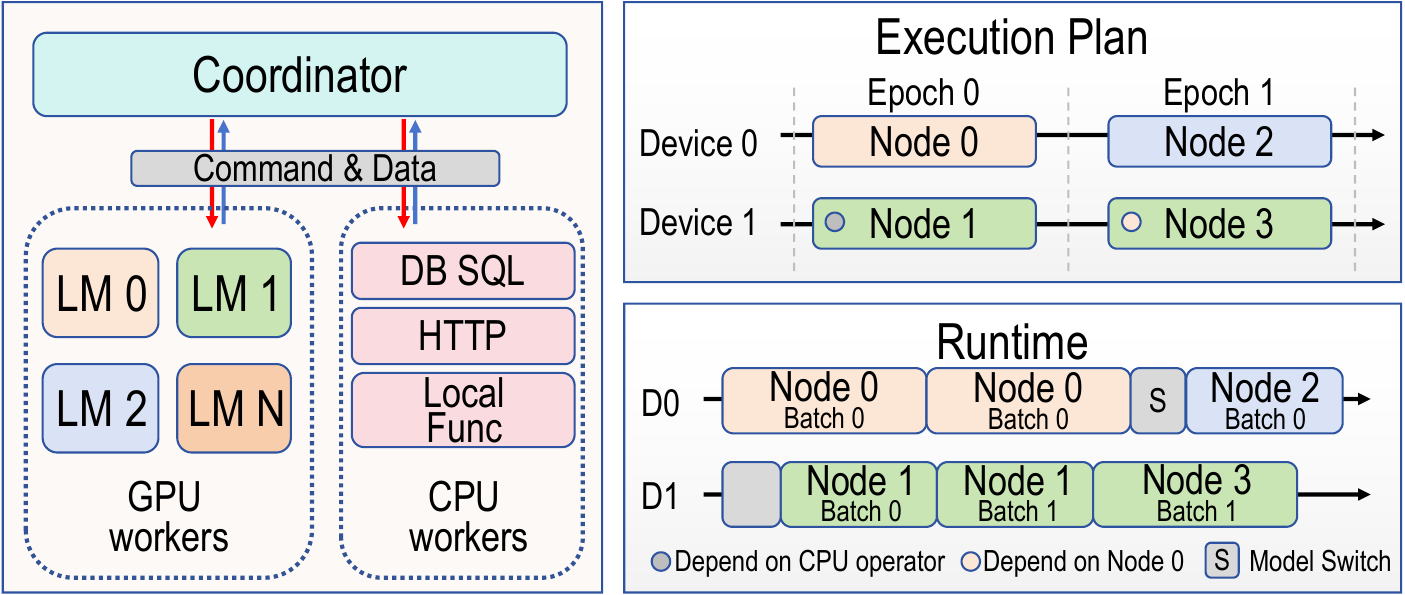}
    \caption{\small{\name's Processor that maps an execution plan to coordinated execution across heterogeneous CPU and GPU workers.}}
    \label{fig:placeholder}
\end{figure}

The Processor materializes the Optimizer's \textsc{ExecutionPlan} into a high-throughput execution over heterogeneous CPU and GPU workers. Its design goal is to preserve the plan's global placement and ordering intent while absorbing runtime stochasticity (e.g., straggling tool calls, decode variance, and network jitter) through asynchronous orchestration and opportunistic overlap.

\smallskip
\noindent\textbf{Coordinator and executors}. The Processor uses a non-blocking, event-driven Coordinator that maintains the global DAG state
(e.g., node readiness, completion, and lineage) and dispatches work according to the epoch plan. The Coordinator runs a lock-light event loop that consumes completion events and promotes newly ready nodes into typed queues.
The runtime manages two classes of long-running workers:
(i) \emph{stateful GPU executors}, each hosting an LLM engine instance (e.g., vLLM-style continuous batching) and maintaining persistent context \(h_w\) (resident weights and warm KV state); and (ii) \emph{generic CPU executors}, which execute tool operators (SQL/HTTP/local functions) and materialize intermediate results.

\smallskip
\noindent\textbf{Dependency resolution and CPU scheduling}. To prevent GPU starvation, the Coordinator prioritizes CPU tasks that unlock the immediate GPU frontier. Concretely, ready tool nodes are ordered by increasing DAG depth to the next unsatisfied LLM node (shallower, more critical prerequisites first). This policy is consistent with the optimizer's cost model, which charges preparation time in \(T_{\text{prep}}\), and it improves CPU--GPU overlap without requiring the solver to explicitly schedule CPU threads.

It is notable that tool execution is CPU- and I/O-bound and can create head-of-line blocking if launched indiscriminately. The Processor enforces bounded concurrency per tool backend (e.g., per-DB connection pool or HTTP client) and applies backpressure when queues exceed configured thresholds, stabilizing latency under bursty workloads.

\smallskip
\noindent\textbf{Work reduction via request coalescing}. Beyond executing tasks, the Processor reduces redundant work across concurrent workflows. For tool operators, it canonicalizes an operator \emph{signature} (operator type plus normalized arguments) and
merges pending tasks with identical signatures into a single physical execution. The resulting output is then fanned out to all dependent logical nodes. This coalescing is particularly effective for high-fanout agents that issue repeated SQL templates or identical retrieval calls.

For structured databases, \name additionally exploits prepared statements within an epoch: it compiles frequently repeated query templates once and reuses the prepared form across merged invocations, reducing parsing/plan overhead under high concurrency.

\begin{table*}[t]
\centering
\small
\setlength{\tabcolsep}{6pt}
\begin{tabular}{lcl >{\raggedright\arraybackslash}p{10cm}}
\toprule
\textbf{ID: Name} & \textbf{\#Nodes} & \textbf{Description} \\
\midrule

\textbf{W1: IMDb-Diamond}
& 8 / 9
& A diamond-shaped topology performing parallel join-heavy SQL queries to stress synchronization. \\

\textbf{W2: IMDb-TripleChain}
& 10 / 3
& Three independent parallel chains with movie, person, and crew data before merging into a final answer. \\

\textbf{W3: FineWiki-LongChain}
& 9 / 6
& A deep sequential chain interleaved with frequent blocking DB retrievals to test critical path optimization. \\

\textbf{W4: FineWiki-Bridge}
& 9 / 3
& A main reasoning chain augmented by periodic auxiliary DB queries, with irregular dependency insertions. \\

\textbf{W5: TPCH-Trident}
& 7 / 9
& A parallel structure launching three concurrent, compute-intensive analytical SQL branches (Q1--Q22). \\

\textbf{W6: TPCH-Fanout}
& 9 / 12
& Broadcasts query parameters to multiple agents across two stages to aggregate complex analytical metrics. \\

\bottomrule
\end{tabular} 
\caption{\small{Evaluation workloads. \#Nodes (GPU/CPU) denotes the count of LLM inference nodes and CPU-bound tool nodes (e.g., SQL execution) respectively.}}\vspace{-0.1in}
\label{tab:workloads}
\end{table*}

\smallskip
\noindent\textbf{GPU execution: batching, chaining, and overlap}. Within each epoch, GPU executors receive the plan's assigned LLM nodes. Each executor feeds requests into its serving engine, which performs continuous batching internally to maximize GPU utilization. The Coordinator preserves the plan's placement decisions to maintain model residency and cache reuse, but does not impose rigid barrier synchronization across workers.
To improve end-to-end latency, the Processor implements wavefront-style execution: when an LLM node finishes, its dependent successors are immediately promoted if their tool prerequisites are satisfied. This enables fast requests to progress downstream without waiting for slower siblings, improving pipeline turnover and reducing fan-in delays.

Indeed, runtime variance can temporarily block planned high-priority tasks (e.g., an external API straggler). To avoid idling, the Coordinator opportunistically executes other ready tasks from its buffer provided that (i) data dependencies are satisfied, and (ii) the execution does not disrupt imminent GPU state (e.g., by forcing a model eviction needed for near-term planned nodes). In practice, this policy acts as a constrained work-stealing mechanism that masks tail latency while preserving the plan's locality intent.

\smallskip
\noindent\textbf{Correctness and plan adherence}. The Processor is semantics-preserving: it never executes a node before its predecessors complete, and coalescing is applied only when operator signatures are identical (ensuring that one physical execution is equivalent to multiple logical ones). When opportunistic execution is enabled, the Coordinator restricts it to actions that do not violate dependencies or undermine the plan's resource guarantees; doing so preserves the Optimizer's global schedule.

\section{Evaluation}
\label{sec:exp}
We now present a comprehensive evaluation of the system performance of \name in comparison with state-of-the-art LLM inference engines. We aim to confirm the following:
\begin{itemize}[leftmargin=0.15in]
    \item G1. {\name} is highly efficient for batch agentic LLM workflows. We showcase wall-clock latency improvements on diverse workloads compared to state-of-the-art LLM and agent serving solutions (Sec.~\ref{sec:exp:e2e}).

    \item G2. The optimizer of {\name} finds near-optimal plans with a low overhead. We verify that our epoch-based DP scheduler generates execution plans matching an offline MILP oracle while reducing planning time by orders of magnitude, and we quantify the critical contributions of our active batching and load guidance mechanisms (Sec.~\ref{sec:optimality}).

    \item G3. {\name} delivers superior throughput for online serving. We demonstrate that our system sustains high query-per-second (QPS) rates under continuous streaming conditions, effectively leveraging pipeline parallelism to outperform baselines in both complex tool-use workflows and linear LLM chains (Sec.~\ref{sec:streaming}).

    \item G4. {\name} exhibits robust scalability and hardware adaptability. We confirm that the system maintains efficiency across varying concurrency levels (scale-invariance), scales linearly with additional worker resources (elasticity), and adapts seamlessly to heterogeneous model sizes and device topologies (Sec.~\ref{sec:ablatio}, \ref{sec:sensitivity-scalability}).
\end{itemize}

\subsection{Setup}

\noindent \textbf{Testbed and models.}
All experiments are conducted on a dedicated server equipped with two AMD EPYC 9755 CPUs, 2.2~TiB of main memory, and three NVIDIA H200~NVL GPUs, each with 141~GiB of HBM.
We evaluate the system using three large language models: Qwen3 (14B and 32B) and GPT-OSS (20B)~\cite{yang2025qwen3, agarwal2025gpt}.

\vspace{0.5em}
\noindent \textbf{Baselines.}
We compare \name against a set of representative LLM serving engines
and agentic workflow systems that cover different points in the design
space, ranging from inference-centric serving to workflow-level
orchestration:

\begin{figure}
    \centering
    \includegraphics[width=1\linewidth]{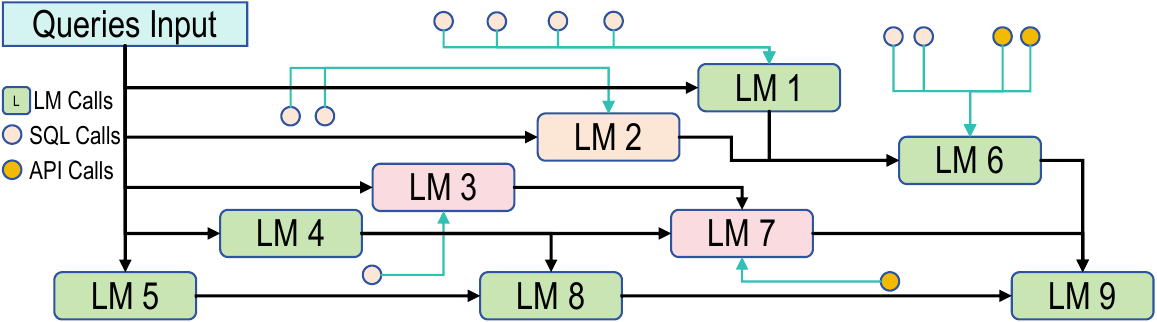}\vspace{-0.05in}
    \caption{\small{Workflow example (W6).}}\vspace{-0.05in}
    \label{fig:w6}
\end{figure}

\begin{itemize}[leftmargin=*]
    \item \emph{vLLM} (Inference-Only): A state-of-the-art inference engine optimizing individual requests via continuous batching~\cite{kwon2023efficient}. We implement agentic workflows in a query-by-query manner: the client driver executes the entire workflow DAG for a single query to completion before initiating the next query. This baseline represents a serial execution mode that isolates the engine's raw latency without the benefits of inter-request batching.

    \item \emph{OpWise} (Topology-Aware Batching): A baseline inspired by the stage-wise execution of MapReduce~\cite{dean2008mapreduce} and Spark~\cite{zaharia2010spark}.
    It utilizes the same batch processor as \name but explicitly maximizes the batch size by buffering all requests at a specific stage before execution.
    However, this strict layer-by-layer synchronization prevents interleaving different operators, leading to resource waste from stragglers.

    \item \emph{LangGraph} (v1.0.4): A popular graph-based orchestration framework~\cite{langgraph}. It represents workflows as explicit DAGs but treats LLM calls as black-box API operations. It schedules tasks based on control flow availability but lacks the system-level feedback to coordinate with the GPU scheduler for efficient placement.

    \item \emph{AgentScope} (v1.0.9): An actor-based multi-agent platform. It encapsulates agents as independent actors communicating via message passing~\cite{gao2402agentscope, gao2025agentscope}. It optimizes the parallel execution of multiple agents to improve throughput. However, the strict actor isolation prevents global state sharing, leading to fragmented execution that cannot be holistically optimized.

    \item \emph{Parrot} (v0.1.0): A prefix-aware serving system that exposes application semantics via "Semantic Variables"~\cite{lin2024parrot}. We follow its best practices by submitting the entire workflow batch upfront, enabling it to perform global dataflow analysis. This represents the state-of-the-art in semantic-aware serving.

    \item \emph{\name}: Our proposed workflow-aware serving system. It features an epoch-based DP scheduler that performs joint optimization of node execution order and worker assignment.
\end{itemize}

All systems use the same \texttt{vLLM} inference backend\cite{kwon2023efficient} and PostgreSQL~\cite{stonebraker1986design} database backend to ensure a fair comparison at the inference and DB-access layers.
All control variables, including sampling parameters and decoding settings, are held constant across systems.
Our prototype implementation of \name integrates vLLM~0.11.0 and SGLang~0.5.6 and PostgreSQL~18.0. For the DP scheduler, we load a Rust core (via PyO3) to accelerate the inner search loops, built once with maturin for the target platform.

\vspace{0.5em}
\noindent \textbf{Workloads and datasets.}
We evaluate \name using a suite of synthetic yet representative multi-stage LLM workflows that capture common structural motifs in production agentic systems.
Each workload is expressed as a DAG combining vLLM-served inference nodes with heterogeneous tool nodes that execute I/O-bound operations, such as SQL queries against a database or HTTP requests to external APIs.
Together, these workflows span a broad range of dependency structures, compute--I/O interleavings, and synchronization patterns.

\vspace{0.5em}
\noindent\emph{Workloads.}
Table~\ref{tab:workloads} summarizes the workloads used in our evaluation.
Each workload is designed to isolate a distinct scheduling challenge commonly observed in real-world agentic pipelines. Figure~\ref{fig:w6} further demonstrates an example workflow.

A key operational difference lies in the deployment model.
Baselines such as \textit{LangGraph}, \textit{AgentScope}, and \textit{Parrot} adopt a separated client-server design that relies on \emph{pre-launched} vLLM backends.
This static requirement obligates users to reserve GPU resources for all model types prior to execution, making it operationally prohibitive to scale beyond a few concurrent models under limited hardware.
In contrast, \name configures execution contexts on the fly, allowing for dynamic model loading and eviction.
To ensure a feasible comparison environment for the baselines, we restrict each workload to utilize at most three distinct model types.
This constraint reflects the rigidity of the baselines' static deployment model rather than a limitation of \name, which naturally accommodates workflows with higher degrees of model heterogeneity.

\begin{figure}[t]
    \centering
    \includegraphics[width=0.95\linewidth]{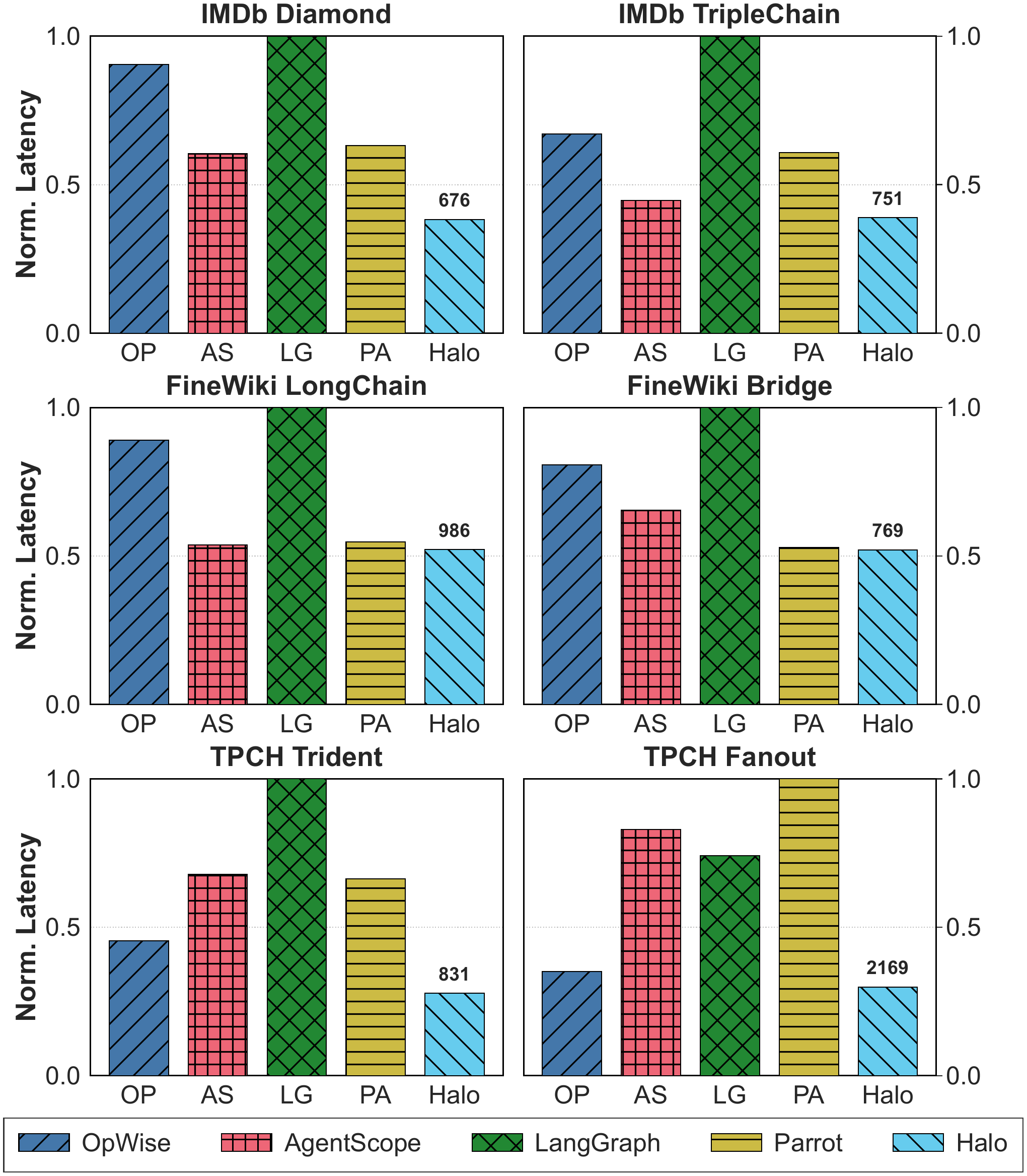}\vspace{-0.1in}
    \caption{\small{End-to-end comparisons of \name with state-of-the-art baselines.} }\vspace{-0.05in}
    \label{fig:e2e_latency}
\end{figure}

\vspace{0.05in}\noindent\emph{Datasets.}
We back all workloads with three relational databases populated from public, widely used datasets to simulate diverse data-intensive agentic tasks.

\begin{itemize}[leftmargin=*]

  \item \emph{FineWiki~\cite{penedo2025finewiki}.}
  We utilize FineWiki (\texttt{HuggingFaceFW/finewiki}), a superior filtered extraction of the Wikipedia~\cite{wikipedia_dump} corpus that retains structural formatting. 
  We ingest a curated subset of {20,000} English page-level records into PostgreSQL, including metadata and raw wikitext.
  To emulate realistic RAG (Retrieval-Augmented Generation) scenarios dominated by point lookups, we create B-tree indexes on primary keys and title columns.

  \item \emph{IMDb~\cite{imdb_interfaces}.}
  For large-scale structured data, we employ the official non-commercial IMDb datasets. 
  The raw data is normalized into a relational schema, resulting in a database with over {200M rows}.
  This dataset captures complex many-to-many relationships between titles, cast, and crew. We extensively index foreign keys to facilitate {multi-way joins}, simulating the heavy relational processing required by knowledge-discovery agents.

  \item \emph{TPC-H~\cite{tpch}.}
  To evaluate analytical and decision-support workloads, we use the TPC-H benchmark generated at scale factor~10 (SF=10). 
  The data is loaded into PostgreSQL following the standard schema, comprising eight core relations with tens of millions of tuples.
  The SQL queries in our workflows are derived from the canonical TPC-H query templates.  Parameters are instantiated dynamically based on the agent's context, ensuring that the workload faithfully reflects standardized analytical patterns while maintaining the variability of agent-driven tool execution.

\end{itemize}

\subsection{End-to-End Benchmarks}
\label{sec:exp:e2e}

We evaluate the end-to-end latency of processing a large batch of input queries ($N=1024$) across the representative agentic workflows.
Figure~\ref{fig:e2e_latency} presents the comparative results.
\name consistently achieves the lowest latency across all datasets, demonstrating the effectiveness of its epoch-based scheduler in balancing batching efficiency with execution fluidity.

\vspace{0.3em}
\noindent\textbf{Comparisons.}
The naive query-by-query execution in \textit{vLLM} results in prohibitive latency for batched agentic workloads. For {W1}, the total runtime for 1024 requests exceeds {71 hours} ($2.56 \times 10^5$ s). In contrast, \name completes the task in just {676 seconds}, achieving a speedup of over {400$\times$}. This extreme gap underscores that without workflow-aware batching, standard inference engines cannot scale to handle complex, multi-turn agentic workloads efficiently.
 
\textit{OpWise} maximizes instantaneous batch size but incurs high model redeployment costs due to its rigid layer-by-layer execution.
In contrast, \name prioritizes LLM engine reuse and CPU-GPU parallelism.
It schedules tasks requiring the same model to execute consecutively, thereby minimizing the frequency of expensive model loading operations.
Additionally, \name overlaps CPU-bound tool execution with GPU inference.
This strategy yields a {2.4$\times$} speedup on {W1} (676s vs. 1595s) and maintains a 1.6$\times$ lead on the complex {W6} workload.
By eliminating the frequent context switching and synchronization barriers inherent to \textit{OpWise}, \name achieves superior throughput.

To compare with the orchestration frameworks, we submit all requests to these frameworks simultaneously to maximize concurrency.
While this allows their backends to leverage engine-level batching for competitive performance on simple workflows (trailing \name by 1.03$\times$--2.6$\times$ on {W1}--{W4}), it relies on \emph{blind execution}.
These decoupled frameworks treat every request independently, missing opportunities for \emph{Request Coalescing} at the workflow level, such as deduplicating redundant tool calls or merging overlapping SQL queries.
Consequently, their performance suffers on complex analytical workloads ({W5}, {W6}) where redundancy is high.
On {W5} and {W6}, \name achieves a {2.4--3.6$\times$} speedup precisely because it systematically identifies and merges redundant I/O operations that the baselines execute individually.
Furthermore, the decoupled design prevents fine-grained memory management during these burst submissions, limiting scalability on multi-model pipelines.

\begin{table}[t]
\centering
\small
\setlength{\tabcolsep}{4pt}
\begin{tabular}{lcccccc} 
\toprule
\multirow{2}{*}{\textbf{Scheduler}} & \multicolumn{3}{c}{\textbf{W1}} & \multicolumn{3}{c}{\textbf{W6}} \\ 
\cmidrule(lr){2-4} \cmidrule(lr){5-7} 
 & \textbf{E2E Lat.} & \textbf{Opt.} & \textbf{Solver} & \textbf{E2E Lat.} & \textbf{Opt.} & \textbf{Solver} \\ 
\midrule
\emph{Random}        & 367.27s & 0.4 & -- & 672.08s & 0.0 & -- \\
\rowcolor{gray!10}
\emph{RR (Ray)}      & 351.17s & 0.4 & -- & 844.46s & 0.0 & -- \\
\emph{Greedy (HEFT)} & 338.28s & 0.6 & -- & 768.18s & 0.2 & -- \\
\rowcolor{gray!10}
\textbf{Halo}        & \textbf{323.64s} & \textbf{1.0} & 2.24s & \textbf{545.55s} & \textbf{1.0} & 1.15s \\
\cmidrule{1-7} 
\emph{MILP (Oracle)} & 322.17s & --  & 1.44h & 544.06s & --  & 172.84s \\
\bottomrule
\end{tabular}
\caption{\small{Comparison of scheduling efficiency. \textbf{Opt.} is the normalized optimality score (see text for details). \textbf{Solver} indicates the computation time required by the scheduling algorithm.}}\vspace{-0.1in}
\label{tab:ablation_w1_w6}
\end{table}

\subsection{Optimality Analysis}
\label{sec:optimality}

We compare four scheduling strategies from recent systems:

\begin{itemize}[leftmargin=*]
\item\emph{Random} dispatches ready operators uniformly at random. It serves as a naive baseline that ignores workflow topology and resource locality.

\item\emph{RayServe-style (RR)} follows the decentralized scheduling principle employed by modern frameworks like Ray~\cite{moritz2018ray}.
It assigns operators to available workers in a simple Round-Robin (RR) manner.
While this strategy respects basic dependencies, it lacks global coordination, often missing opportunities to optimize for data locality and downstream resource contention.

\item\emph{HEFT} represents the classic greedy, list-based DAG scheduling algorithm~\cite{topcuoglu2002performance}.
It prioritizes operators using local heuristics and greedily maps them to resources that minimize the estimated cost.
Similarly, recent ML-specific systems like \emph{JellyBean}~\cite{wu2023jellybean} also rely on greedy heuristics for model selection and placement on heterogeneous infrastructures. While these approaches account for locality, the greedy nature limits their view to local optima.

\item\emph{\name} employs our epoch-based dynamic programming scheduler, which jointly optimizes operator ordering, batching, and placement under a unified global objective.
\end{itemize}

\vspace{0.05in}\noindent\textbf{Oracle and comparisons.}
We implement an Oracle planner inspired by the compilation strategy of Alpa~\cite{zheng2022alpa, li2023alpaserve}.
Specifically, we formulate a continuous-time \emph{Mixed-Integer Linear Programming (MILP)} model to derive the global optimum ($S^{\star}$). Besides, we define a GPU-focused optimality metric, $\text{Opt}(S)$, to measure how closely a schedule matches the oracle's decisions:
\[
\text{Opt}(S)=\max_{\pi}\;
\frac{\bigl|\mathcal{P}(S)\cap \pi(\mathcal{P}(S^{\star}))\bigr|}
{\bigl|\mathcal{P}(S^{\star})\bigr|},
\]
where $\mathcal{P}(\cdot)$ denotes the set of ordered pairs of {consecutive nodes} executing on the same GPU.
This metric quantifies the recall of optimal execution dependencies preserved by the scheduler.

While $S^{\star}$ provides the theoretical ceiling for performance, solving the MILP is computationally prohibitive.
For example, on {W1}, the solver requires over {1.44 hours} ($5.19 \times 10^6$\,ms) to converge.
In contrast, \name computes a near-optimal schedule in just 2.24 seconds, yielding a $2{,}322\times$ speedup in planning time while matching the oracle's quality.

\vspace{0.05in}\noindent\textbf{Analysis.}
To evaluate scheduler optimality under varying conditions, we employ two distinct workloads, {W1} and {W6} (N = 256). Table~\ref{tab:ablation_w1_w6} reveals a strong correlation between scheduling optimality and normalized end-to-end workload latency. The performance gap between baselines and the optimal solution significantly widens as the workload topology becomes more complex.

We found that \emph{Random} and \emph{RR (Ray)} have large overheads, with a $1.55\times$ latency on {W6}. Their topology-agnostic assignment fails to account for data movement and model switching costs, resulting in execution plans with low optimality (0.0--0.4). Greedy \emph{(HEFT)} performs adequately on the simpler {W1} ($1.05\times$) but degrades significantly on the complex {W6} ($1.41\times$). This result highlights the critical and often conflicting trade-off between model reuse and pipeline parallelism. Greedy heuristics typically over-prioritize immediate locality to avoid switching costs. However, in complex topologies like W6, this local optimization tends to inadvertently serialize independent branches or overload specific workers, leading to resource underutilization and missed opportunities for concurrency.

In contrast, \name consistently achieves near optimality across both workloads.
Moreover, our DP-based schedule exhibits a {high degree of alignment} with the MILP Oracle, producing execution plans that are structurally identical to the theoretical optimum. This alignment validates the {rationality} of our discretization algorithm.

\begin{figure}[t]
\centering
\includegraphics[width=0.7\linewidth]{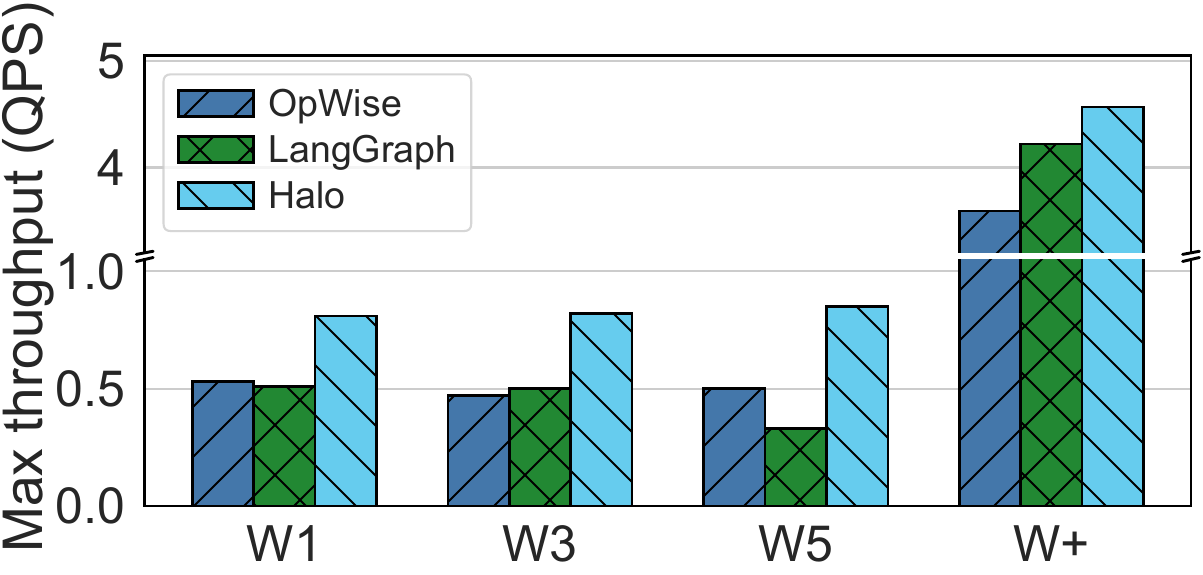}\vspace{-0.1in}
\caption{\small{Online throughput comparison. \name consistently outperforms \textit{OpWise} and \textit{LangGraph} across different workloads.}}
\label{fig:streaming-throughput}
\end{figure}

\subsection{Online Serving Performance}
\label{sec:streaming}

To evaluate \name in a realistic online serving scenario, we subject the system to a continuous stream of asynchronous requests.
Requests are ingested into an input queue and dispatched to the execution engine to measure the maximum sustainable throughput (QPS). Other experiment settings remain the same as in Sec.~\ref{sec:exp:e2e}.
We benchmark against \emph{OpWise} and the agentic framework \emph{LangGraph} to the performance advantages of \name over both state-of-the-art optimizations and industry-standard tools.
In addition to the standard benchmarks (W1, W3, W5), we also leverage {W+} (a lightweight, LLM-only linear chain with 3 nodes) to assess performance in rapid-response scenarios without CPU-bound tool overheads.

Figure~\ref{fig:streaming-throughput} demonstrates that \name consistently delivers superior throughput over both baselines.
On complex workflows ({W1--W5}), \name achieves speedups ranging from {1.53$\times$} to {1.74$\times$} over \emph{OpWise} (e.g., 0.82 vs. 0.47 QPS on {W3}).
This performance gap stems from \name's superior model scheduling and execution parallelism, which minimize idle time compared to \emph{OpWise}'s rigid execution.
Against \emph{LangGraph}, the advantage peaks at {2.58$\times$} on the TPCH workflow ({W5}, 0.85 vs. 0.33 QPS).
This gain is driven by the synergy of \name's \emph{Request Coalescing}, which eliminates redundancy in batch processing, and its \emph{global scheduler}, which optimizes resource pipelining under high concurrency.
Even on the lightweight {W+} workload, \name (4.57 QPS) leads \emph{OpWise} (3.59 QPS) and \emph{LangGraph} (4.22 QPS), proving that our architectural optimizations remain effective even for simple, pure-inference chains.

\subsection{Ablation Study}
\label{sec:ablatio}

We conduct an ablation study on both {W1} and {W6} workloads to quantify the individual contributions of \name's scheduling and optimization mechanisms.
Table~\ref{tab:component-ablation} reports the latency changes relative to the full \name system. Below we discuss the findings of the experiments. 

\begin{table}[t]
\centering
\small
\setlength{\tabcolsep}{10pt}
\begin{tabular}{lcc}
\toprule
\multirow{2}{*}{\textbf{Ablation Variant}} & \textbf{W1} & \textbf{W6} \\
\cmidrule(lr){2-2} \cmidrule(lr){3-3}
 & \textbf{E2E Latency}  & \textbf{E2E Latency}  \\
\midrule
\emph{w/o Profiling Scoring}       & {+20\%} & +8\% \\
\rowcolor{gray!10}
\emph{w/o CPU Load Guidance}       & +1\% & {+18\%} \\
\emph{w/o Opportunistic Exec.}     & +1\% & {+56\%} \\
\rowcolor{gray!10}
\emph{w/o Request Coalescing}      & +16\% & {+154\%} \\
\cmidrule{1-3}
\textbf{\name (Full)}              & {1.00} & {1.00} \\
\bottomrule
\end{tabular}
\caption{\small{Component ablation analysis on \textbf{W1} and \textbf{W6}. Higher values indicate larger performance drops when the component is removed.}}  
\label{tab:component-ablation}
\end{table}

First, replacing the historical profile-based scoring with a naive heuristic based on dependency in a {1.20$\times$ latency increase} on {W1}.
The naive approach inaccurately estimates node costs by equating upstream dependency \emph{quantity} with computational \emph{intensity}.
Consequently, the scheduler misidentifies lightweight aggregation nodes as expensive tasks, deprioritizing them behind heavy computations.
This static misjudgment disrupts optimal execution ordering, creating avoidable pipeline stalls.

Next, disabling CPU load guidance ($T_{prep}$) leads to a {1.18$\times$ latency increase} on {W6}.
This mechanism is critical for ensuring {intra-epoch load balancing} between CPU and GPU resources. Lacking explicit visibility into CPU pressure, the scheduler tends to saturate workers with locality-preferred tasks while leaving GPUs idle, creating significant resource skew within a single scheduling cycle.
\name leverages this guidance to modulate task dispatching, ensuring that CPU processing and GPU inference remain balanced within each epoch to maximize pipeline throughput.

Removing the opportunistic execution mechanism leads to a {1.56$\times$} slowdown on complex topologies ({W6}). This feature empowers the backend Processor to {proactively identify} and execute ready nodes during idle cycles (e.g., I/O stalls), provided they do not violate the global dependency order.
Without this adaptive capability, the Processor is strictly bound to the scheduler's static dispatch rate, leaving available CPU/GPU cycles unutilized during minor scheduling gaps or data transfer latencies.

Lastly, the \emph{w/o Request Coalescing} variant exhibits the most severe performance degradation, causing a {2.54$\times$} slowdown on {W6}.
Unlike naive systems that blindly process every generated request individually, \name employs a workflow-level coalescing strategy.
It scans all queries within the current batch to identify and merge redundant I/O operations (e.g., identical SQL queries generated by high-fanout agents).
Disabling this module forces the backend to process the raw, unreduced stream of requests, exposing the system to the full complexity of the workload and causing severe I/O congestion.

\subsection{Sensitivity and Scalability Analysis}
\label{sec:sensitivity-scalability}

For simplicity, we evaluate the robustness and adaptability of \name using {W3} workflow variants with input size 256. We employ \textit{OpWise} as a standard data-system baseline to benchmark how well the system maintains efficiency across dynamic environments and varying scales.

\begin{figure}[t]
    \centering
    \includegraphics[width=\linewidth]{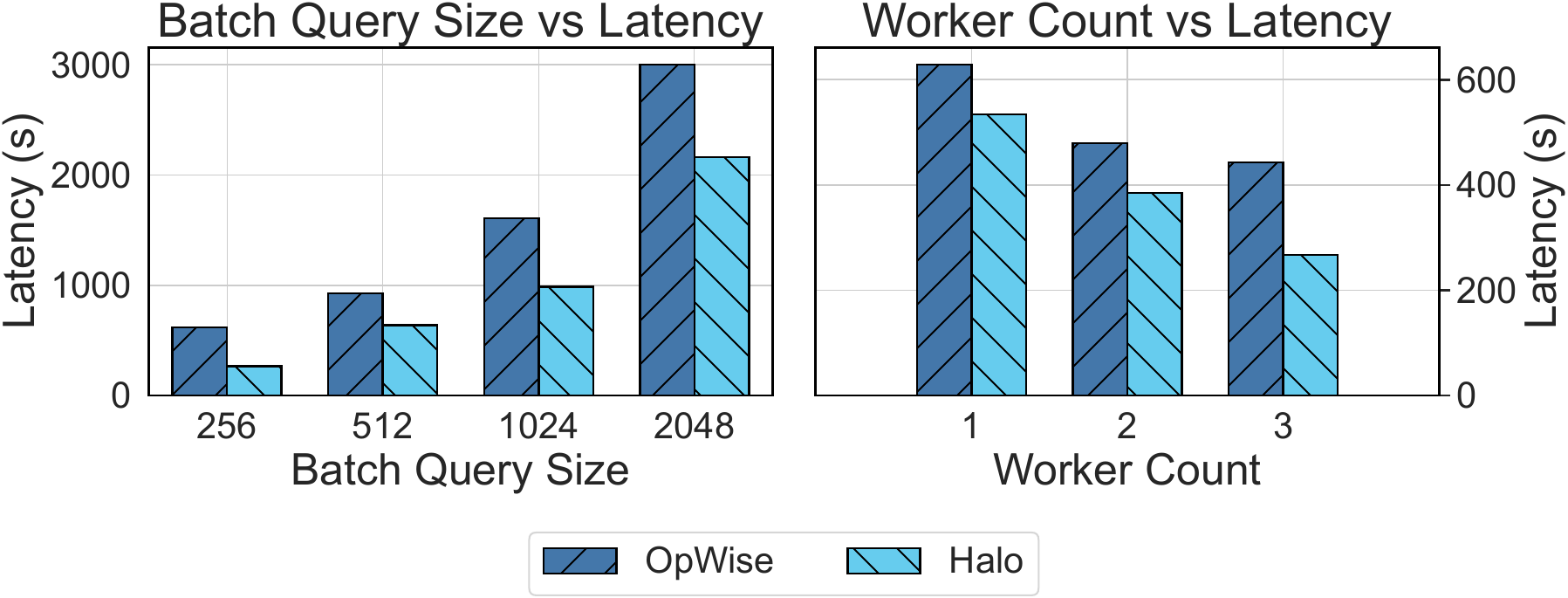}\vspace{-0.1in}
    \caption{\small{Scalability analysis of \name.}}
    \label{fig:scalability}
\end{figure}
\begin{figure}[t]
    \centering
    \includegraphics[width=\linewidth]{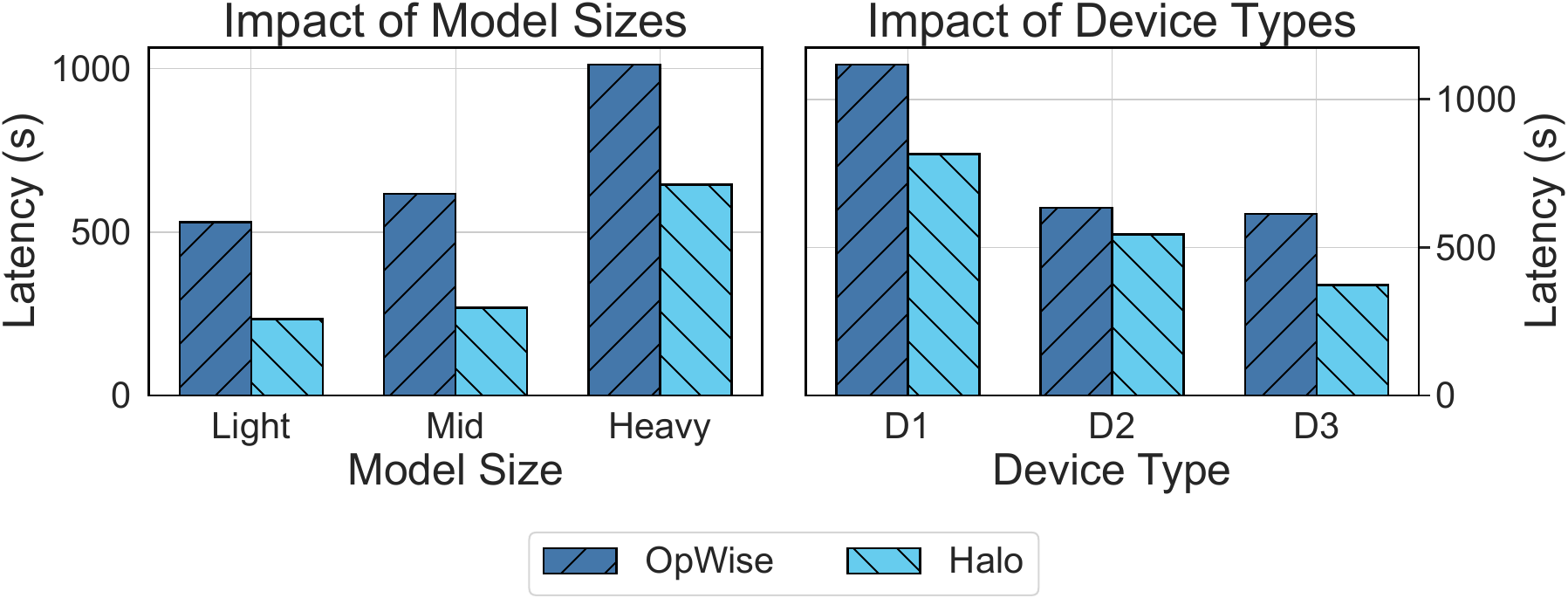}\vspace{-0.1in}
    \caption{\small{Sensitivity analysis of \name. D1-D3 refer to different GPUs.}}  
    \label{fig:sensitivity}
\end{figure}

\vspace{0.05in}\noindent\textbf{Scalability analysis.}
We evaluate \name by stretching on the degree of parallelism and scaling workers. Figure~\ref{fig:scalability} shows our findings.

\begin{itemize}[leftmargin=*]
\item\textit{Impact of batch query size.}
We vary the input batch query size from 256 to 2048. \name demonstrates near-linear stability across these scales. We note that the efficiency of \emph{OpWise} improves with a larger input batch size, but this happens only when the batch size reaches 2048, which is practically high and risking OOM errors. \name keeps its efficiency advantage across different input batch sizes. 

\item\textit{Impact of worker count.}
We examine \name's elasticity using 1 to 3 H200 GPU workers.
The results confirm that \name possesses a good elasticity with a near-linear performance gain. The base \emph{OpWise} faces a diminishing performance gain due to synchronization constraints. The queuing and job dispatching mechanism in \name's scheduler ensures scalability with elastic resources.
\end{itemize}

\begin{figure}[t]
    \centering
    \includegraphics[width=\linewidth]{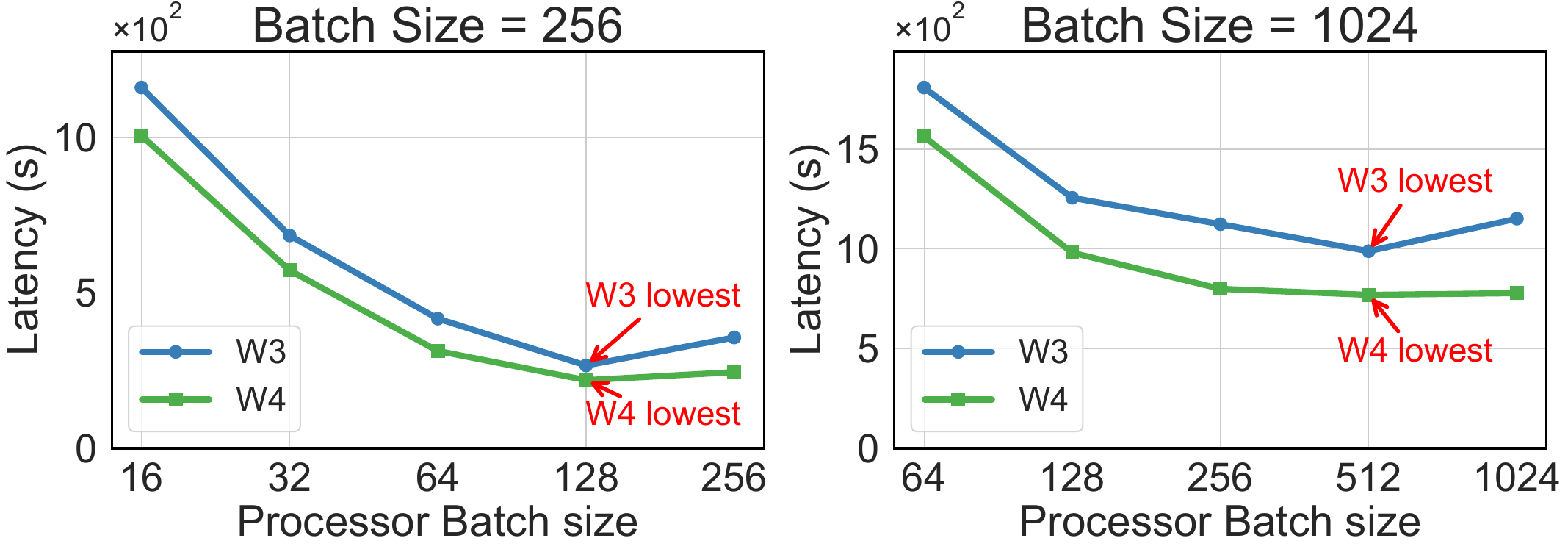}\vspace{-0.1in}
    \caption{\small{Sensitivity analysis with different Processor batch sizes.}}
    \label{fig:pp-batch-sensitivity}
\end{figure}

\vspace{0.05in}\noindent\textbf{Sensitivity analysis.}
We assess the robustness of \name by verifying its adaptation to diverse model complexities and heterogeneous hardware topologies (Figure~\ref{fig:sensitivity}), as well as different LLM batch inference sizes (Figure~\ref{fig:pp-batch-sensitivity}).

\begin{itemize}[leftmargin=*]
\item\textit{Impact of model size.}
We deploy models ranging from \textit{Light} (0.4B--4B)~\cite{yang2025qwen3} to \textit{Heavy} (32B variants)~\cite{guo2025deepseek, team2025qwq} to test adaptability. Results show that \name remains {model-agnostic}.
In the \textit{Light} setting, \name suppresses overhead induced by scheduling decisions to match rapid inference speeds; in the \textit{Heavy} setting, it seamlessly shifts to maximizing compute utilization.

\item\textit{Impact of device type.}
We evaluate \name's robustness with different GPU workers: {D1} (2$\times$A100), {D2} (2$\times$H100), and {D3} (3$\times$H200).
The results highlight \name on heterogeneous hardware to offer gains on all setups compared with the \emph{OpWise} baseline. 

\item\textit{Impact of processor batch size.}
Recall that \name's Processor employs its own batch size during execution, i.e., batch LLM inference or SQL queries that are processed at a time. We conduct a sensitivity analysis on {W3} and {W4} under moderate (256) and heavy (1024) query loads. Figure~\ref{fig:pp-batch-sensitivity} identifies that processor batch size affects system efficiency. Latency drops initially, driven by the synergy of hardware and data optimization to saturate workers and amortize overhead. However, performance gains diminish: excessive batching prolongs stage execution, which stifles pipeline fluidity by delaying inter-stage data propagation and introducing {head-of-line blocking} that starves downstream stages. Consequently, \name targets the inflection point that maximizes hardware utilization without compromising rotational frequency.

\end{itemize}

\subsection{Case Study}
\label{sec:case-study}

\begin{figure}[t]
    \centering
    \includegraphics[width=\linewidth]{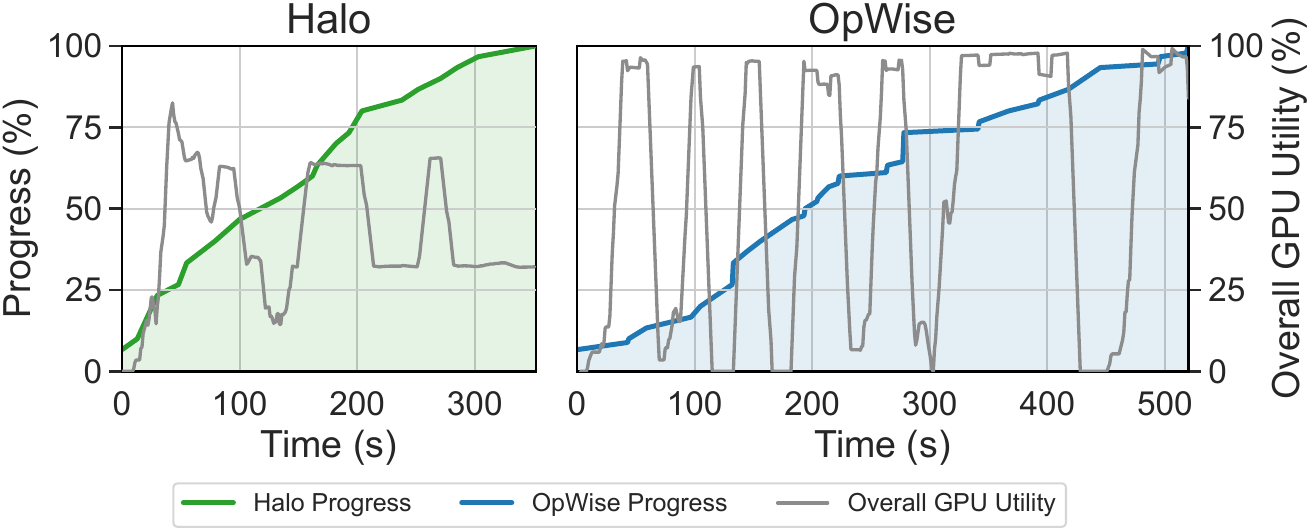}\vspace{-0.1in}
    \caption{\small{Real-time execution progress and GPU utilization on W3 (256 inputs). \name demonstrates faster completion with a stable resource profile, compared to the bursty and inefficient locking patterns of \textit{OpWise}.}}
    \label{fig:case_study}
\end{figure}

To provide a deeper insight into the execution dynamics, we profile the runtime behavior of the {W3} workload with 256 query inputs. Figure~\ref{fig:case_study} visualizes and compares the real-time execution progress (colored lines) and the instantaneous GPU utilization (gray lines) for \name and \textit{OpWise}.

As shown in the progress trajectories, \name demonstrates a steeper and more consistent slope compared to \textit{OpWise}. This accelerated completion rate stems from \name's higher degree of parallelism; its non-blocking scheduler effectively overlaps GPU-bound inference with I/O-bound tasks (e.g., DB queries), thereby masking latency and preventing the "stop-and-wait" delays that retard \textit{OpWise}.
The GPU utilization profiles further corroborate this efficiency. Following the analysis methodology in cluster scheduling~\cite{xiao2020antman}, we observe that \textit{OpWise} exhibits a bursty, "sawtooth" pattern, oscillating between saturation (near 100\%) and idleness. These fluctuations indicate inefficient resource locking and data movement. In contrast, \name maintains a steadier execution with fewer idle gaps. 

To quantify the cost implications, we calculate the {Cumulative GPU Usage} ($\int_{0}^{T} U(t) dt$), which serves as a proxy for the total {GPU-seconds} consumed—the primary billing unit in cloud environments~\cite{yang2023skypilot}. \textit{OpWise} accumulates a substantial resource footprint of {28,889} unit$\cdot$s. Conversely, \name completes the workload significantly earlier with a much leaner footprint of {14,690} unit$\cdot$s. This {$\approx$2.0$\times$ reduction in GPU-seconds} highlights that \name achieves high throughput not by brute-forcing utilization, but by eliminating wasteful resource holding.

\subsection{Limitations}
\label{sec:limitations}
Our experiments demonstrate the effectiveness of \name. Nevertheless, we are still working towards a broader scope and a more production-ready solution. We note the following limitations: 

\begin{itemize}[leftmargin=*]
  \item \emph{Deployment scope.} Halo targets a single-machine, multi-GPU setting to enable white-box control over GPU memory, batching, and locality; extending the same guarantees to multi-node deployments would require additional mechanisms for distributed placement, cache movement, and network-aware scheduling.
  \item \emph{Fixed-plan with semantics-preserving optimization.} Halo assumes a fixed logical workflow DAG and enforces strict semantic equivalence (no prompt rewriting or approximate techniques), which limits opportunities for higher-level rewrites that trade accuracy for cost or adapt the workflow structure online.
\end{itemize}
\section{Related Work}
\label{sec:related_work}

\noindent \textbf{LLM serving and inference optimization.} 
The rapid deployment of LLMs has driven significant innovation in inference systems. Early optimizations focused on memory management and continuous batching, with \emph{vLLM}~\cite{kwon2023efficient} introducing \textit{PagedAttention} to eliminate KV-cache fragmentation. To address the memory wall in long-context scenarios, \emph{SGLang}~\cite{zheng2024sglang} proposed \textit{RadixAttention} for automatic prefix sharing, while systems like \emph{CacheGen}~\cite{liu2024cachegen} and \emph{LMCache}~\cite{cheng2025lmcache} explored disaggregated storage for latency reduction. 
More recently, disaggregated architectures like \emph{DistServe}~\cite{zhong2024distserve} and \emph{Splitwise}~\cite{patel2024splitwise} have isolated prefill and decode phases to resolve resource interference. 
However, these systems primarily optimize at the engine or request level. They lack the global visibility required to perform plan-level optimizations—such as node reordering or operator fusion—across complex agentic DAGs, which \name{} addresses.

\vspace{0.05in}\noindent \textbf{Multi-agent frameworks.} 
Frameworks such as \emph{LangGraph}~\cite{langgraph}, \emph{AutoGen}~\cite{wu2024autogen}, and \emph{AgentScope}~\cite{gao2025agentscope} simplify the logical construction of agentic applications by providing abstractions for state management and message passing. While these frameworks excel at defining agent behaviors, they typically treat the underlying LLM inference as a black-box API. This decoupling prevents the runtime from exploiting structural dependencies within the workflow, leading to redundant computations and suboptimal hardware utilization compared to \name{}'s white-box optimization approach.

\vspace{0.05in}\noindent \textbf{System support for agentic workflows.}
Recent work has begun to bridge the gap between frameworks and engines. 
\emph{Ayo} (previously \emph{Teola})~\cite{tan2025ayo} introduces a fine-grained orchestration framework that compiles workflows into primitive-based dataflow graphs, enabling primitive-level pipelining. 
Similarly, \emph{Parrot}~\cite{lin2024parrot} facilitates KV-reuse through semantic variables. 
Both primarily target single-query latency. In contrast, \name{} optimizes batch throughput via a unified cost model. 
Additionally, while \emph{KVFlow}~\cite{pan2025kvflow} improves memory efficiency with workflow-aware eviction, it remains a reactive policy; \name{} instead proactively integrates cache states into global scheduling.
More recently, \emph{Murakkab}~\cite{chaudhry2025murakkab} proposes a declarative, SLO-aware serving stack for multi-tenant agentic workflows that decouples workflow specification from execution configuration, and uses profiling plus an adaptive runtime to select models/hardware and reconfigure execution to meet SLOs. 
In contrast, \name{} assumes a fixed logical plan and focuses on physical intra-batch optimization—CPU--GPU co-scheduling, operator reordering, and batching—leveraging DP-style planning to maximize throughput for dependency-rich DAG execution.

\vspace{0.05in}\noindent \textbf{Heterogeneous scheduling and pipelining.}
Prior systems exploit hardware heterogeneity either at the cluster level or via pipeline parallelism.
Cluster schedulers such as \emph{Gavel}~\cite{narayanan2020heterogeneity} (and recent follow-ups~\cite{sultana2025resource}) allocate heterogeneous accelerators across deep-learning jobs using throughput-based models, optimizing makespan/JCT and utilization.
Execution-level pipeline systems (e.g., \emph{PipeDream}~\cite{narayanan2019pipedream} and \emph{HetPipe}~\cite{park2020hetpipe}) optimize stage partitioning/mapping for training on non-uniform GPUs, often using DP-style planners to reduce pipeline bubbles.
In contrast, \name{} targets \emph{inference-time} \emph{agentic DAGs} that mix GPU LLM operators with CPU tool operators (SQL/HTTP/parsing), and performs \emph{plan-level} intra-batch optimization (placement, reordering, batching) to balance CPU--GPU load and exploit cache/tool states.
Methodologically, \name{} inherits the DP spirit from pipeline planning but generalizes it from linear pipelines to dependency-rich, multi-model DAG scheduling.

\vspace{0.05in}\noindent \textbf{Query optimization for AI.} 
\name{} draws inspiration from database query optimization. Systems like \emph{DocETL}~\cite{shankar2024docetl}, \emph{Palimpzest}~\cite{liu2025palimpzest}, and \emph{LOTUS}~\cite{patel2025semantic} employ declarative languages and cost-based optimizers to select efficient logical plans or rewrite pipelines for accuracy and cost. 
Crucially, these systems focus on \textit{logical} restructuring (e.g., model selection or prompt rewriting). In contrast, \name{} assumes a fixed logical plan and focuses on \textit{physical} execution optimization (scheduling, placement, batching) on heterogeneous hardware. \name{} can thus be viewed as the efficient runtime engine that underpins these high-level logical optimizers.
\section{Conclusion}
\label{sec6}
We presented \name, a workflow-aware query processor for agentic LLM workloads. \name models each request as a DAG spanning GPU LLM operators and CPU tool operators, and applies cost-based scheduling and placement to overlap execution, preserve locality, and exploit cross-request batching and sharing. Experiments show that \name improves end-to-end latency and online throughput over inference-only serving and existing orchestration frameworks with low planning overhead. We believe \name provides a practical runtime substrate for scalable agentic analytics.


\bibliographystyle{ACM-Reference-Format}
\bibliography{references}

\appendix

\end{document}